\documentclass[twocolumn,showpacs,preprintnumbers,amsmath,amssymb]{revtex4}

\usepackage{graphicx}% Include figure files
\usepackage{dcolumn}% Align table columns on decimal point
\usepackage{bm}% bold math

\begin{document}

\title{Superconductor coupled to two Luttinger liquids\\
as an entangler for electron spins}
\author{Patrik Recher and Daniel Loss}
\address{ Department of Physics and Astronomy, University of
Basel,\\ Klingelbergstrasse 82, CH-4056 Basel, Switzerland}

\date{December 16, 2001}

\begin{abstract}
We consider an s-wave superconductor (SC) which is tunnel-coupled
to two spatially separated Luttinger liquid (LL) leads. We
demonstrate that such a setup acts as an entangler, i.e. it
creates spin-singlets of two electrons which are spatially
separated, thereby providing a source of electronic
Einstein-Podolsky-Rosen pairs. We show that in the presence of a
bias voltage, which is smaller than the energy gap in the SC,
 a stationary current of spin-entangled electrons can flow from the
SC to the LL leads due to Andreev tunneling events. We discuss two
competing transport channels for Cooper pairs to tunnel from the
SC into the LL leads. On the one hand,  the coherent tunneling of
two electrons into the same LL lead is shown to be suppressed by
strong LL correlations  compared to single-electron tunneling into
a LL. On the other hand, the tunneling of two spin-entangled
electrons into different leads is suppressed by the initial
spatial separation of the two electrons coming from the same
Cooper pair. We show that the latter suppression depends crucially
on the effective dimensionality of the SC. We identify a regime of
experimental interest in which the separation of two
spin-entangled electrons is favored. We determine the decay of the
singlet state of two electrons injected into different leads
caused by the LL correlations. Although the electron is not a
proper quasiparticle of the LL, the spin information can still be
transported via the spin density fluctuations produced by the
injected spin-entangled electrons.
\end{abstract}

\pacs{73.40.Gk, 71.10.P, 74.10.P}

%\vskip2pc]
\narrowtext
\maketitle
\section{Introduction}
Pairwise and non-local entangled quantum states, so-called
Einstein-Podolsky-Rosen (EPR) pairs \cite{Einstein}, represent the
fundamental resource for quantum communication \cite{Bennett84}
schemes like dense coding, quantum teleportation or  quantum key
distribution \cite{BennettNature}, or more fundamentally, they can
be used to test Bell's inequalities \cite{Bell}. Experiments have
tested Bell's inequalities \cite{Aspect}, dense coding
\cite{Mattle}, and quantum teleportation \cite{Bouwmeester,Boschi}
using photons, but to date no experiments for {\em massive }
particles like electrons in a solid state environment exist.  This
is so because it is difficult to first produce entangled electrons
and also to detect them afterwards in a controlled way due to
other electrons interacting with the entangled pair. On the other
hand, the spin of an electron was pointed out to be a most natural
candidate for a quantum bit (qubit) \cite{Loss97,QCReview}. This
idea was supported also by experiments which show unusually long
dephasing times for electron spins in semiconductors (approaching
microseconds) and phase coherent transport up to $100\:\mu{\rm m}$
\cite{Kikkawa1,Kikkawa2,Awschalom}. In addition, the electron also
possesses charge which makes it well suited for transporting the
spin information \cite{MMM2000,BLS}. Further, the Coulomb
interaction between the electron charges can be exploited to
spatially separate the spin-entangled electrons resulting in
electronic EPR pairs. As first pointed out in Refs.
\cite{MMM2000,BLS}, such electronic EPR pairs can be used for
testing Bell inequalities  and for quantum communication schemes
in the solid state. The first step towards this goal is to have a
scheme by which the electrons can be reliably entangled. One
possibility is to use coupled quantum dots \cite{MMM2000,BLS}.
Alternatively, we recently proposed an entangler device
\cite{RSL}, which creates mobile and non-local spin-entangled
electrons, consisting of an s-wave superconductor (SC), where the
electrons are correlated in Cooper pairs with spin-singlet
wavefunctions \cite{note}. The SC is tunnel-coupled via two
quantum dots in the Coulomb blockage regime \cite{Kouwenhoven} to
two spatially separated Fermi liquid leads. By applying a bias, a
stationary current of spin-entangled electrons can flow from the
SC to the leads. The quantum dots are used to mediate the
necessary interaction between the two electrons initially forming
a Cooper pair in the SC so that the two electrons tunnel
preferably not into the same but instead into different leads.
This entangler then satisfies all requirements to detect the
entanglement via the current noise in a beam splitter setup
\cite{BLS}. It is straightforward to formulate spin measurements
for testing Bell inequalities (it is most promising to measure
spin via charge \cite{Loss97,Recher,EL}).
 We refer to related work \cite{Lesovik,Falci,Melin}, which makes also use of
Andreev tunneling, but in a regime opposite to the one considered
 in Ref. \cite{RSL} and here, where the superconductor/normal interface is
 transparent and no Coulomb blockade nor strong
correlations are present.

In the present work we propose and discuss an alternative
realization of an entangler which is based on strongly interacting
one-dimensional wires which show Luttinger liquid (LL) behavior.
In comparison to our earlier proposal with quantum dots
\cite{RSL}, we replace now the Coulomb blockade behavior of the
dots by strong correlations of the LL. Well-known examples for LL
candidates are carbon nanotubes \cite{Bockrath}. The low energy
excitations of these LL are collective charge and spin modes
rather than quasiparticles which resemble free electrons like they
exist in a Fermi liquid. As a consequence, the single-electron
tunneling into a LL is suppressed by strong correlations.  The
question then arises quite naturally whether these strong
correlations can even further suppress the coherent tunneling of
{\em two} electrons into the same LL, as provided by a correlated
two-particle tunneling event (Andreev tunneling), so that the two
electrons preferably separate and tunnel into different LL leads.
It turns out that the answer is positive. To address this question
we introduce a setup consisting of an s-wave SC which is weakly
tunnel-coupled to the center (bulk) of two spatially separated
one-dimensional wires 1,2 described as Luttinger liquids, see
Fig.\ 1,2. In this model we calculate the stationary current
generated by the tunneling of a singlet (spin-entangled
electrons), transferred from the SC into two separate leads
(non-local process) or into the same  lead (local process), 1 or
2. We show that the ratio of these two competing current channels
depends on the system parameters and that it can be made large in
order to have the desired injection of the two electrons in two
separate leads, where, again, the two spins, forming a singlet,
are entangled in spin space while separated in orbital space and
therefore represent an electronic EPR pair.
\begin{figure}[h]
\centerline{\includegraphics[width=6cm]{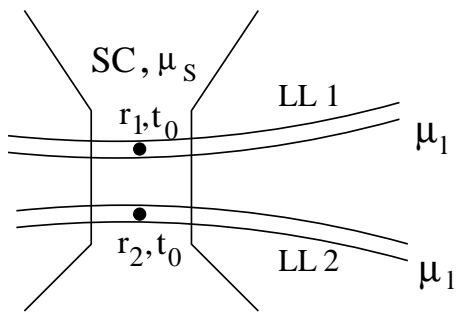} \vspace{1mm}}
\label{fig1} \caption{A possible implementation of the entangler
setup: Two quantum wires 1,2, described as infinitely long
Luttinger liquids (LL), are deposited on top of an s-wave
superconductor (SC) with chemical potential $\mu_{S}$. The
electrons of a Cooper pair can tunnel by means of an Andreev
process from two points ${\bf r}_{1}$ and ${\bf r}_{2}$ on the SC
to the center (bulk) of the two quantum wires 1 and 2 resp. with
tunneling amplitude $t_{0}$. The interaction between the leads is
assumed to be negligible.}
\end{figure}
It is well-known that tunneling of single electrons into LLs is
suppressed compared to Fermi liquids due to strong many-body
correlation. In addition, we find now that subsequent tunneling of
a second electron into the same LL is further suppressed, again in
a characteristic interaction dependent power law, provided the
applied voltage bias between the SC and the LL is much smaller
than the energy gap $\Delta$ in the SC so that single-electron
tunneling is suppressed. The two-particle tunneling event is
strongly correlated within the uncertainty time $\hbar/\Delta$,
characterizing the time-delay between subsequent tunneling events
of the two electrons of the same Cooper pair. In other words, the
second electron of a Cooper pair is incluenced  by the existence
of its preceding partner electron already present in the LL. This
effect can also be interpreted as a Coulomb blockade effect,
similar to what occurs in quantum dots attached to a SC
\cite{CBL,RSL}. Similar Coulomb blockade  effects occur also in a
mesoscopic chiral LL within a quantum dot coupled to macroscopic
chiral LL edge-states in the fractional quantum Hall regime
\cite{GL}. There, the Coulomb blockade-like energy gap is
quantized in units of the non-interacting energy level spacing of
the quantum dot and its existence is therefore a finite size
effect, whereas in the present case, we will see that the
suppression comes from strong correlations in a two-particle
tunneling event which is  present even  in an infinitely long  LL
as considered here. On the other hand, if the two electrons of a
Cooper pair tunnel to different leads, they will preferably tunnel
from different points from the SC ${\bf r}_{1}$ and ${\bf r}_{2}$,
with distance $\delta {\bf r}= {\bf r}_{1}-{\bf r}_{2}$ due to the
spatial separation of the leads, see Fig.\ 1,2. We find that the
current is exponentially suppressed if the distance $\delta r$
exceeds the coherence length $\xi$ of a Cooper pair on the SC.
This limitation poses no severe experimental restriction since
$\xi$ is on the order of micrometers for usual s-wave materials,
and $\delta r$ can be assumed to be on the order of nanometers.
Still, a power law suppression $\propto 1/(k_{F}\delta r)^{2}$,
with $k_{F}$ being the Fermi wavevector in the SC, remains and is
more relevant.
\begin{figure}[h]
\centerline{\includegraphics[width=5cm]{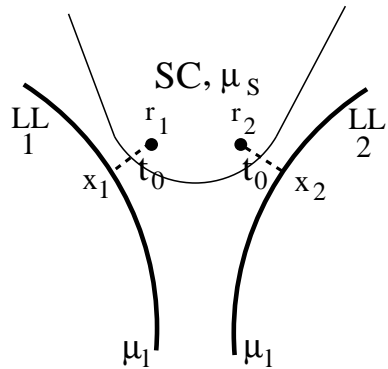} \vspace{1mm}}
\label{fig2} \caption{An alternative implementation of the
proposed entangler set-up: The two quantum wires 1,2 with chemical
potential $\mu_{l}$, described as  infinitely long Luttinger
liquids (LL),  are tunnel-coupled with amplitude $t_{0}$ from two
points $x_{1}$ and $x_{2}$ to two points ${\bf r}_{1}$ and ${\bf
r}_{2}$ of a superconducting (SC) tip with chemical potential
$\mu_{S}$. }
\end{figure}
We show, however, that in lower dimensions of the SC this
suppression is less pronounced (with smaller powers). Further, we
then discuss the decay in time of a spin-singlet state injected
into two LL, one electron in each lead, due to the interaction
present in the LL and find a characteristic power law decay in
time at zero temperature. Despite this  decay of the singlet
state, the spin information can still be transported through the
LL wires via the spin-density fluctuations created by the injected
electrons.

\section {Model and Hamiltonian}
We consider an s-wave  superconductor (SC) which is weakly tunnel-coupled
to the center (bulk) of two spatially separated  Luttinger liquid
(LL) leads (see Fig.\ 1,2). The Hamiltonian of the whole system is
represented as $H=H_{0}+H_{T}$ with
$H_{0}=H_{S}+\sum_{n=1,2}H_{Ln}$ describing the isolated SC and
LL-leads 1,2 respectively. Tunneling between the SC and the leads
is governed by the tunneling Hamiltonian $H_{T}$. Each part of the
system will be described in the following.

The s-wave SC  with chemical potential $\mu_{S}$ is described by
the BCS-Hamiltonian\cite{Schrieffer}

\begin{equation}
H_{S}-\mu_{S}N_{S}=\sum\limits_{{\bf k},s}E_{\bf k}  \gamma_{{\bf
k}s}^{\dagger} \gamma_{{\bf k}s},
\end{equation}
where $s=(\uparrow,\downarrow)$ and $N_{S}=\sum_{{\bf k}s}c_{{\bf
k}s}^{\dagger}c_{{\bf k}s}$ is the number operator for electrons
in the SC. The quasiparticle operators $\gamma_{{\bf k}s}$
describe excitations out of the BCS-groundstate $|0\rangle_{S}$
defined by $\gamma_{{\bf k}s}|0\rangle_{S}=0$.  They are related
to the electron annihilation and creation operators $c_{{\bf k}s}$
and $c_{{\bf k}s}^{\dagger}$ through the Bogoliubov transformation
\cite{Schrieffer}
\begin{eqnarray}
c_{{\bf k}\uparrow}&=&u_{\bf k}\gamma_{{\bf k}\uparrow}+v_{\bf
k}\gamma_{-{\bf k}\downarrow}^{\dagger}\nonumber\\ c_{-{\bf
k}\downarrow}&=&u_{\bf k}\gamma_{-{\bf k}\downarrow}- v_{\bf
k}\gamma_{{\bf k}\uparrow}^{\dagger}\, , \label{Bogoliubov}
\end{eqnarray}
where $u_{\bf k}=(1/\sqrt{2})(1+\xi_{\bf k}/E_{\bf k})^{1/2}$ and
$ v_{\bf k}=(1/\sqrt{2})(1-\xi_{\bf k}/E_{\bf k})^{1/2}$ are the
usual BCS coherence factors \cite{Schrieffer}, and $\xi_{\bf
k}=\epsilon_{\bf k}-\mu_{S}$ is the normal state single-electron
energy counted from the Fermi level $\mu_{S}$, and $E_{\bf
k}=\sqrt{\xi_{\bf k}^2+\Delta^2}$ is the quasiparticle energy. The
field operator for an electron with spin $s$ is $\Psi_{s}({\bf
r})=V^{-1/2}\sum_{{\bf k}}e^{i{\bf kr}}c_{{\bf k}s}$, where $V$ is
the volume of the SC.\\ The two leads $1,2$ are supposed to be
infinite one dimensional interacting electron systems described by
LL-theory. We only include forward scattering
processes which describe scattering of electrons on the same
branch (left or right movers). We neglect backscattering
interactions which involve large momentum transfers of order
$2p_{F}$ where $p_{F}$ is the Fermi wavevector in the LL
\cite{Kanenote}. The LL-Hamiltonian for the low energy excitations
of lead $n=1,2$ can then be written in a bosonized form as
\cite{Schulz}

\begin{eqnarray}
H_{Ln}-\mu_{l}N_{n}=\qquad\qquad&&\nonumber\\
\sum_{\nu=\rho,\sigma}\int\limits_{-L/2}^{+L/2}dx\,\left
(\frac{\pi u_{\nu}K_{\nu}}{2}\,\Pi_{n\nu}^2+ \frac{u_{\nu}}{2\pi
K_{\nu}}\,(\partial_{x}\phi_{n\nu})^2\right),&& \label{LL
Hamiltonian}
\end{eqnarray}

where the fields $\Pi_{n}(x)$ and $\phi_{n}(x)$ satisfy bosonic
commutation relations $\left [\phi_{n\nu}(x),\Pi_{m\mu}(x')\right
]=i\delta_{nm}\delta_{\nu\mu}\delta(x-x')$, and $\mu_{l}$ is the
chemical potential of the LL-leads (assumed to be identical for
both leads), and $N_{n}=\sum_{s}\int
dx\,\psi_{ns}^{\dagger}(x)\psi_{ns}(x)$ is the number operator for
electrons in LL $n$. The Hamiltonian (\ref{LL Hamiltonian})
describes long-wavelength  charge ($\nu=\rho$) and spin
($\nu=\sigma$) density oscillations in the LL propagating with
velocities $u_{\rho}$ and $u_{\sigma}$, respectively. The
velocities $u_{\nu}$ and the stiffness parameters $K_{\nu}$ depend
on the interactions between the electrons in the LL.  In the limit
of vanishing backscattering, we have $u_{\sigma}=v_{F}$ and
$K_{\sigma}=1$, and the LL is described by only two parameters
$K_{\rho}<1$ and $u_{\rho}$. In a system with full translational
invariance we have $u_{\rho}=v_{F}/K_{\rho}$. We decompose the
field operator describing electrons with spin $s$ into a right and
left moving part,
$\psi_{ns}(x)=e^{ip_{F}x}\psi_{ns+}(x)+e^{-ip_{F}x}\psi_{ns-}(x)$.
The right (left) moving field operator $\psi_{ns+}(x)$
$(\psi_{ns-}(x))$  is then expressed as an exponential of bosonic
fields as \cite{Haldane,Heidenreich}

\begin{eqnarray}
\psi_{ns\pm}(x)&=\lim\limits_{\alpha\to
0}&\frac{\eta_{\pm,ns}}{\sqrt{2\pi\alpha}}\exp\Big\{\pm\frac{i}{\sqrt{2}}\Big(\phi_{n\rho}(x)\Big.\Big.\nonumber\\
&&\Big.\Big.\quad+s\phi_{n\sigma}(x)\mp(\theta_{n\rho}(x)+s\theta_{n\sigma}(x))\Big)\Big\},\nonumber\\
&&
\label{bosonization}
\end{eqnarray}
where $[\phi_{n\nu}(x),\theta_{m\mu}(x')]=-i(\pi/2)\delta_{nm}\delta_{\nu\mu}{\rm sgn}(x-x')$ and therefore $\partial_{x}\theta_{n\nu}(x)=\pi\,\Pi_{n\nu}(x)$. The
operators $\eta_{\pm,ns}$ are needed to ensure the correct
fermionic anticommutation relations.  In the thermodynamic (TD)
limit ($L\rightarrow\infty$), $\eta_{\pm,ns}$ can be presented by
Hermitian operators  satisfying  the anticommutation relation
\cite{Schulz} $\{\eta_{r},\eta_{r'}\}=2\delta_{rr'}$, with
$r=\pm,ns$. We adopt the convention throughout the paper that
$s=+1$ for $s=\uparrow$, and $s=-1$ for $s=\downarrow$, if $s$ has
not the meaning of an operator index.\\ Transfer of electrons from
the SC to the LL-leads is described by the tunneling Hamiltonian
$H_{T}=\sum_{n}H_{Tn}+{\rm H.c.}$, where $H_{Tn}$ is defined as
$H_{Tn}=t_{0}\,\sum_{s}\,\psi_{ns}^{\dagger}\Psi_{s}({\bf
r}_{n})$.  The field operator $\Psi_{s}({\bf r}_{n})$ annihilates
an electron with spin $s$ at point ${\bf r}_{n}$ on the SC,  and
$\psi_{ns}^{\dagger}$ creates it again with amplitude $t_{0}$ at
point $x_{n}$ in the LL $n$ which is nearest to ${\bf r}_{n}$, see
Fig.\ 1,2. We assume that the spin is conserved during the
tunneling process, and thus the tunneling amplitudes $t_{0}$ do
not depend on spin, and, for simplicity, are the same for both
leads $n=1,2$. We remark that our point-contact approach for
describing the electron transfer from the SC to the LL is the
simplest possible description but it captures presumably the
relevant features of a real device. The scheme shown in Fig.\ 2
has a geometry which suggests that electrons tunnel from point
${\bf r}_{n}\rightarrow x_{n}$ which are closest to each other,
due to the fact that $t_{0}$ depends exponentially on the
tunneling distance. In the setup shown in Fig.\ 1, a point-like
tunnel contact between the SC and the LL might be induced by
slightly bending the quantum wires (e.g. nanotubes). If the
contact area has a finite extension, we note that the two
electrons preferably tunnel from the same point on the SC, when
they tunnel into the same lead, since the two-particle tunneling
event is coherent and shows a suppression in the probability
already on a length scale given by $1/k_{F}$, as we discuss in
detail below.

\section{ Stationary current from the SC to the LL-leads}
%\hspace*{1cm}Two competing transport channels.}
We now calculate the current of singlets, i.e. pairwise
spin-entangled electrons (Cooper pairs), from the SC to the
LL-leads due to Andreev tunneling \cite{Glazman,Fazio} in first
non-vanishing order, starting from a general T-matrix approach
\cite{Merzbacher}. We thereby distinguish two transport channels.
First we calculate the current when two electrons tunnel from
different points ${\bf r}_{1}$ and ${\bf r}_{2}$ of the SC into
{\em different} interacting LL-leads which are separated in space
such that there is no inter-lead interaction. In this case the
only correlation in the tunneling process is due to the
superconducting pairing of electrons which results in a coherent
two-electron tunneling process of opposite spins from different
points ${\bf r}_{1}$ and ${\bf r}_{2}$ of the SC, and with a delay
time $\sim \hbar/\Delta$ between the two tunneling events. Since
the total spin is a conserved quantity $[H,{\bf S}^2]=0$, the spin
entanglement of a Cooper pair is transported to {\em different}
LL-leads, thus leading to non-local spin-entanglement. On the
other hand, if two electrons tunnel from the same point of the SC
into the {\em same} LL-lead there is an additional correlation in
the LL-lead itself due to the intra-lead interaction. It is the
goal of this work to investigate how the transport current for
tunneling of two electrons from the SC into the {\em same }
LL-lead is effected by this additional correlation.\\

\section{The T-matrix }
\label{T-matrix} We apply a T-matrix (transmission matrix)
approach \cite{Merzbacher} to calculate the current \cite{RSL}.
The stationary current of {\em two } electrons passing from the SC
to the leads is then given by
\begin{equation}
I=2e\sum_{f,i}W_{fi} \rho_{i}.
\end{equation}
Here, $W_{fi}$  is the transition rate from  the superconductor to
the leads, given by
$
W_{fi}=2\pi |\langle
f|T(\varepsilon_i)|i\rangle|^2\delta(\varepsilon_f-\varepsilon_i)\,
.
$
Here, $T(\varepsilon_i)=H_{T}\frac{1}{\varepsilon_i +i
\eta-H}(\varepsilon_i-H_{0})$ is the on-shell transmission or
T-matrix, with $\eta$ being a positive infinitesimal which we set
to zero at the end of the calculation. The T-matrix can be
expanded in a power series in the tunneling  Hamiltonian $H_{T}$,
\begin{equation}
\label{series} T(\varepsilon_i)=H_{T}+H_{T}\sum_{n=1}^{\infty}
(\frac{1}{\varepsilon_i+i\eta-H_{0}}H_{T})^n \, ,
\end{equation}
where $\varepsilon_{i}$ is the energy of the initial state
$|i\rangle$, which, in our case, is the energy of a Cooper pair at
the Fermi surface of the SC, $\varepsilon_{i}=2\mu_{S}$. Finally,
$\rho_{i}=\langle i|\rho|i\rangle$ is the stationary occupation
probability for the entire system to be in the state $|i\rangle$.
We work in the regime $\Delta>\mu>k_{B}T$, where
$\mu=\mu_{S}-\mu_{l}$ is the applied voltage bias between the SC
and the leads, and $T$ the temperature with $k_{B}$ the Boltzmann
constant. The regime $\Delta>\mu$ ensures that single electron
tunneling from the SC to the leads is excluded and only tunneling
of {\em two} coherent electrons of opposite spins is allowed. In
the regime $\mu>k_{B}T$ we only have transport from the SC to the
leads, and not in the opposite direction. Since temperature is
assumed to be the smallest energy scale in the system, we assume
$k_{B}T=0$ in the calculation.  The set of initial states
$|i\rangle$, virtual states $|v\rangle$ and final states
$|f\rangle$ consists of the BCS groundstate (GS) $|0\rangle_{S}$
and excitations $\gamma_{{\bf k}s}^{\dagger}|0\rangle_{S}$ for the
SC and a complete set \cite{Haldane,Heidenreich} of energy
eigenstates $|N_{nr\nu},\{b_{n\nu}\}\rangle$ of the LL-Hamiltonian
$H_{Ln}$ given in (\ref{LL Hamiltonian}). $N_{nr\nu}$ is the
number of excess spin ($\nu=\sigma$) and charge ($\nu=\rho$)  in
branch $r$ relative to the state where all single-particle states
are filled up to the chemical potential $\mu_{l}$. The Bose
operators $b_{n\nu}$ form a continuous spectrum describing
collective spin and charge modes and will be introduced in (\ref{bogo2}) and (\ref{bogo3}).  The GS of the
LL is then $|0,{0}\rangle$, which means that we have a filled
Fermi sea and no bosonic excitations. The energy contribution of
the excess charge and spin is included in the so-called zero mode
($k=0$) terms in the diagonalized Hamiltonian $K_{Ln}$
(\ref{diag}) and are of no importance in the TD-limit
($L\rightarrow \infty$) considered here, since the contribution of
these terms due to an additional electron on top of the GS is
${\cal O}(1/L)$ and is neglected in (\ref{diag}). For a detailed
description of the LL-Hamiltonian (\ref{LL Hamiltonian}) including
the zero-modes see Appendix A. Since we want to calculate the
transition rate for transport of a Cooper pair to the leads, the
final states $|f\rangle$ of interest contain two additional
electrons of opposite spins in the leads compared to the initial
state $|i\rangle$.\\
\\
\section{ current $I_{1}$ for tunneling of two electrons into different leads}
\label{current1S} We first calculate the current for tunneling of
two spin-entangled electrons into different leads. We expand the
T-matrix to second order in $H_{T}$ and go over to the interaction
representation by using $\delta
(\epsilon)=(1/2\pi)\int_{-\infty}^{+\infty}dt\,e^{i\epsilon t}$,
and $ \langle
v|(\epsilon_{i}-H_{0}+i\eta)^{-1}|v\rangle=-i\int_{0}^{\infty}dt\,e^{i(\epsilon_{i}-\epsilon_{v}+i\eta)t}$.
By transforming the time dependent phases into a time dependence
of the tunneling Hamiltonian we can integrate out all final and
virtual states. The forward current $I_{1}$ for tunneling of two
electrons into different leads can then be written as

\begin{eqnarray}
I_{1}=2e\lim_{\eta \to 0}\sum_{n\neq n'\atop m\neq
m'}\,\int\limits_{-\infty}^{\infty}dt\int\limits_{0}^{\infty}dt'\int\limits_{0}^{\infty}dt''\,e^{-\eta(t'+t'')}&&\nonumber\\
e^{i(2t-t'-t'')\mu}\,\langle
H_{Tm}^{\dagger}(t-t'')H_{Tm'}^{\dagger}(t)H_{Tn}(t')H_{Tn'}\rangle\,,&&
\label{current1}
\end{eqnarray}
where $\langle\cdots\rangle$ denotes $\rm{ Tr}\rho\{\cdots\}$. The
bias has been introduced in a standard way \cite{Mahan}, and the
time dependence of the operators in (\ref{current1}) is then
governed by
$H_{Tn}(t)=e^{i(K_{Ln}+K_{S})t}H_{Tn}e^{-i(K_{Ln}+K_{S})t}$ with
$K_{Ln}+K_{S}=H_{Ln}+H_{S}-\mu_{l}N_{n}-\mu_{S}N_{S}$. The
transport process involves two electrons of {\em different} spins
which suggests that the average in (\ref{current1}) is of the form
(suppressing time variables),
$\langle\cdots\rangle=\sum_{ss'}\langle
H_{Tm-s'}^{\dagger}H_{Tm's'}^{\dagger}H_{Tn-s}H_{Tn's}\rangle$,
where $H_{Tns}$ describes tunneling of spin $s$ governed by
$H_{Tn}$. The time sequence in (\ref{current1}) contains the
dynamics of the hopping of a Cooper pair from the SC to the
LL-leads (one electron per lead) and back. The times $t'$ and
$t''$ are delay times between subsequent hoppings of two electrons
from the {\em same} Cooper pair, whereas $t$ is the time between
injecting and taking out a Cooper pair. We evaluate the thermal
average in (\ref{current1}) at zero temperature where the
expectation value is to be taken in the groundstate of $K_{0}$
which is the  BCS groundstate of the SC and the bosonic vacuum of
the LL-leads for a filled Fermi sea. We remark that since the
interaction between the different subsystems ($SC, L_{1}, L_{2}$)
is included in the tunneling-perturbation, the expectation value
factorizes into a SC-part times a LL-part. In addition, the
LL-correlation function factorizes into two single-particle
correlation functions due to the negligible interaction between
the LL-leads $1,2$ (this will be  not the case if two electrons
tunnel into the {\em same} lead). Note that in the TD-limit the
time dynamics of all LL-correlation functions will be goverened by
a Hamiltonian that  depends only on Bose operators (see
(\ref{diag})). The operators $\eta_{\pm,rs}$ commute with all Bose
operators, and as a consequence  $\eta_{\pm,rs}$ are time
independent. Therefore, interaction terms of the LL of the form
$\psi_{\alpha}\psi_{\beta}\psi_{\gamma}^{\dagger}\psi_{\delta}^{\dagger}$
can be written as
$\eta_{\alpha}\eta_{\beta}\eta_{\gamma}\eta_{\delta}\times$(Bose
operators), where $\alpha,\beta,\gamma,\delta$ are composite
indices containing $r=\pm,ns$. The correlation function in
(\ref{current1}) is then of the form

\begin{eqnarray}
&&\sum_{n\neq n'\atop m\neq m'}\,\langle
H_{Tm}^{\dagger}(t-t'')H_{Tm'}^{\dagger}(t)H_{Tn}(t')H_{Tn'}\rangle\nonumber\\
\nonumber\\ &&=|t_{0}|^4\,\sum\limits_{s\atop n\neq
m}\big\langle\psi_{ns}(t-t'')\psi_{ns}^{\dagger}\big\rangle\big\langle\psi_{m-s}(t-t')\psi_{m-s}^{\dagger}\big\rangle\nonumber\\
&&\times\big\langle\Psi_{s}^{\dagger}({\bf
r}_{n},t-t'')\Psi_{-s}^{\dagger}({\bf r}_{m},t)\Psi_{-s}({\bf
r}_{m},t')\Psi_{s}({\bf r}_{n})\big\rangle\nonumber\\
&&\nonumber\\ &&-|t_{0}|^4\,\sum\limits_{s\atop n\neq
m}\big\langle\psi_{m-s}(t-t'-t'')\psi_{m-s}^{\dagger}\big\rangle\big\langle\psi_{ns}(t)\psi_{ns}^{\dagger}\big\rangle\nonumber\\
&&\times\big\langle\Psi_{-s}^{\dagger}({\bf
r}_{m},t-t'')\Psi_{s}^{\dagger}({\bf r}_{n},t)\Psi_{-s}({\bf
r}_{m},t')\Psi_{s}({\bf r}_{n})\big\rangle\,.
\label{totcorrelation}
\end{eqnarray}
The 4-point correlation functions of the SC can be calculated by
Fourier decomposing
$
\Psi_{s}({\bf r}_{n},t)=V^{-1/2}\\\sum_{{\bf k}}(u_{{\bf
k}s}\gamma_{{\bf k}s}e^{-iE_{{\bf k}}t}+v_{{\bf k}s}\gamma_{-{\bf
k}-s}^{\dagger}e^{iE_{\bf k}t})e^{i{\bf k}{\bf r}_{n}} $, with
$u_{{\bf k}s}=u_{{\bf k}}$, and $v_{{\bf k}\uparrow}=-v_{{\bf
k}\downarrow}=v_{\bf k}$. For the first correlation function in
(\ref{totcorrelation}) we then obtain
\begin{eqnarray}
\label{Andreev1} &&V^{2}\,\big\langle\Psi_{s}^{\dagger}({\bf
r}_{n},t-t'')\Psi_{-s}^{\dagger}({\bf r}_{m},t)\Psi_{-s}({\bf
r}_{m},t')\Psi_{s}({\bf r}_{n})\big\rangle\nonumber\\
&&\nonumber\\ &&=\sum\limits_{{\bf k}{{\bf k}'}}\,u_{\bf k}v_{\bf
k}u_{{\bf k}'}v_{{\bf k}'}\,e^{-iE_{\bf k}t'}\,e^{iE_{{\bf
k}'}t''}e^{i({\bf k}+{\bf k}')\delta{\bf r}}\nonumber\\
&&+\sum\limits_{{\bf k}{{\bf k}'}}\,\big(v_{\bf k}v_{{\bf
k}'}\big)^{2}\,e^{-iE_{\bf k}(t-t'')}\,e^{-iE_{{\bf k}'}(t-t')},
\end{eqnarray}
where $\delta{\bf r}={\bf r}_{1}-{\bf r}_{2}$ is the distance
vector between the two tunneling points in the SC. The first sum
in (\ref{Andreev1}) describes the (time-dependent) correlation of
creating and annihilating a quasiparticle  (with same spin),
whereas the second term in (\ref{Andreev1}) describes correlation
of creating {\em two} quasiparticles (with different spin). It is
obvious that the second term describes processes which involve
final states $|f\rangle$ in the T-matrix element $\langle
f|T(\varepsilon_i)|i\rangle$ that contain two excitations in the
SC and, therefore, does not describe an Andreev process. In the
regime $\Delta>\mu$ such a process is not allowed by energy
conservation. We will see this explicitly by calculating the
integral over $t$ which originates from the Fourier representation
of the $\delta$-function present in the rate $W_{fi}$. Similarly,
for the correlator $\langle\Psi_{-s}^{\dagger}({\bf
r}_{m},t-t'')\Psi_{s}^{\dagger}({\bf r}_{n},t)\Psi_{-s}({\bf
r}_{m},t')\Psi_{s}({\bf r}_{n})\rangle$ in (\ref{totcorrelation})
we obtain  (\ref{Andreev1}) with a minus sign, and we have to
replace $t-t''$ by $t-t'-t''$, and $t-t'$ by $t$, in the second
term of (\ref{Andreev1}).

To evaluate the LL-correlation functions in (\ref{totcorrelation})
we decompose the phase fields $\phi_{n\nu}(x,t)$ and
$\theta_{n\nu}(x,t)$ into a sum over the spin and charge bosons
(see also Appendix A),

\begin{eqnarray}
\theta_{n\nu}(x,t)&=&-\sum\limits_{p}\,{\rm
sgn}(p)\sqrt{\frac{\pi}{2LK_{\nu}|p|}}\,e^{ipx}\,e^{-\alpha|p|/2}\nonumber
\\ &&\times\big(b_{n\nu
p}\,e^{-iu_{\nu}|p|t}-b_{n\nu-p}^{\dagger}\,e^{iu_{\nu}|p|t}\big)\,,
\label{bogo2}
\end{eqnarray}
and
\begin{eqnarray}
\phi_{n\nu}(x,t)&=&\sum\limits_{p}\,\sqrt{\frac{\pi
K_{\nu}}{2L|p|}}\,e^{ipx}\,e^{-\alpha|p|/2}\nonumber\\
&&\times\big(b_{n\nu
p}\,e^{-iu_{\nu}|p|t}+b_{n\nu-p}^{\dagger}\,e^{iu_{\nu}|p|t}\big).
\label{bogo3}
\end{eqnarray}
The spin and charge bosons satisfy Bose-commutation relations, in
particular  $[b_{n\nu p},b_{n'\nu'p'}^{\dagger}]=\delta_{rr'}$,
where $r\equiv n\nu p$, and the LL-groundstate is defined as
$b_{n\nu p}|0\rangle_{LL}=0$. The Hamiltonian (\ref{LL
Hamiltonian}) can then be written in terms of the $b$-operators as
(see Appendix A)
\begin{equation}
K_{Ln}=\sum_{\nu p}u_{\nu}|p|\,b_{n\nu p}^{\dagger}b_{n\nu p}\,,
\label{diag}
\end{equation}
where we have subtracted the zero-point energy coming from the
filled Dirac sea of negative-energy particle states. In all
$p$-sums  we will explicitly exclude $p=0$ as discussed in Section
\ref{T-matrix} and is explained in more detail in Appendix A. To
account for the p-dependence of the interaction, we apply a high
momentum-transfer cut-off $\Lambda$ on the order of $1/p_{F}$ so
that $K_{\nu}(p)=K_{\nu},\,u_{\nu}(p)=u_{\nu}$ for $|p|<1/\Lambda$
and $K_{\nu}(p)=1,\, u_{\nu}(p)=v_{F}$ for $|p|>1/\Lambda$. By
writing
$\psi_{nsr}(x,t)=(2\pi\alpha)^{-1/2}\eta_{r,ns}\,e^{i\Phi_{nsr}(x,t)}$
with $r=\pm$ and  $\Phi_{nsr}$ defined according to
(\ref{bosonization}), we can represent the single-particle
LL-correlation function as
$\langle\psi_{nsr}(x,t)\psi_{nsr}^{\dagger}\rangle=(2\pi\alpha)^{-1}\exp\{\langle\Phi_{nsr}(x,t)\Phi_{nsr}-(\Phi_{nsr}^{2}(x,t)+\Phi_{nrs}^{2})/2\rangle\}$
with the well-known result \cite{Solyom,Suzumura,Schulz2,Voit}

\begin{eqnarray}
&&G_{nrs}^{1}(x,t)\equiv\left\langle\psi_{nsr}(x,t)\psi_{nsr}^{\dagger}\right\rangle\nonumber\\
&&\nonumber\\ &&=\frac{1}{2\pi}\,\lim\limits_{\alpha\to
0}\frac{\Lambda+i(v_{F}t-rx)}{\alpha+i(v_{F}t-rx)}\nonumber\\
&&\times\prod_{\nu=\rho,\sigma}\,\frac{1}{\sqrt{\Lambda+i(u_{\nu}t-rx)}}\,\left[\frac{\Lambda^{2}}{\left(\Lambda+iu_{\nu}t\right)^{2}+x^{2}}\right]^{\gamma_{\nu}/2}
\label{Greensfunction}
\end{eqnarray}
where $\gamma_{\nu}=(K_{\nu}+K_{\nu}^{-1})/4-1/2>0$ is an
interaction dependent parameter which describes the power law
decay of the long time and long distance correlations. The factor
in the first line of (\ref{Greensfunction}) is only important if
one is interested in $x,t$ satisfying $|v_{F}t-rx|<\Lambda$ and is
a result of including the p-dependence of the interaction
parameters $K_{\nu}$ and $u_{\nu}$. The LL-correlation function
(\ref{Greensfunction}) has singular points as a function of time
in the upper complex plane. It is now clear that the  integration
over $t$ is only nonzero if the phase of $e^{i\omega t}$ in
(\ref{current1}) is positive, i.e. $\omega >0$. Since the phases
of the terms  containing $(v_{\bf k})^2$ in (\ref{Andreev1}) depend
also on $t$, the requirement for a nonzero contribution to the
current from these  terms requires $2\mu-E_{\bf k}-E_{\bf k'}>0$.
This, however, is excluded in our regime of interest, since
$E_{\bf k}+E_{{\bf k}'}\ge2\Delta$, and therefore, we will not
consider these  terms  any further in what follows. We remark that for
$r\neq r'$, the correlation function
$\langle\psi_{nsr}(x,t)\psi_{nsr'}^{\dagger}\rangle$ gives a
negligible contribution in the TD-limit. This statement is true
also for a finite size LL as long as the interaction preserves the
total number of right- and left-movers. In our model the electrons
tunnel into the same point $x_{n}$ in LL $n$, i.e. $x=0$ in
(\ref{Greensfunction}). In addition,  LL 1 and 2 are assumed to be
similar. We therefore have $G_{nrs}^{1}(x,t)\equiv G^{1}(t)$, and
the current $I_{1}$ can then be written as
\begin{eqnarray}
&&I_{1}=(32e\,|t_{0}|^{4}/V^{2})\nonumber\\ &&\times\lim_{\eta\to
0}
\int\limits_{-\infty}^{\infty}dt\int\limits_{0}^{\infty}dt'\int\limits_{0}^{\infty}dt''\,e^{-\eta(t'+t'')}
e^{i(2t-t'-t'')\mu}\,\nonumber\\ &&\times\,\sum\limits_{{\bf
k}{{\bf k}'}}\,u_{\bf k}v_{\bf k}u_{{\bf k}'}v_{{\bf
k}'}\,e^{-iE_{\bf k}t'}\,e^{iE_{{\bf k}'}t''}e^{i({\bf k}+{\bf
k}')\delta{\bf r}}\nonumber\\
&&\times\,\left\{G^{1}(t-t'')G^{1}(t-t')+G^{1}(t-t'-t'')G^{1}(t)\right\}.\nonumber\\
&&
\label{current2}
\end{eqnarray}

We evaluate (\ref{current2}) in leading order in the small
parameter $\mu/\Delta$ and remark that the delay times $t'$ and
$t''$ are restricted to $t',t''\stackrel{<}{\sim}1/\Delta$. This
becomes clear if we set $\delta {\bf r}=0$ and express the
contribution in  (\ref{current2}) containing the dynamics of the
SC as
\begin{eqnarray}
&&\sum_{{\bf k}{{\bf k}'}}\,u_{\bf k}v_{\bf k}u_{{\bf k}'}v_{{\bf
k}'}\,e^{-iE_{\bf k}t'}\,e^{iE_{{\bf k}'}t''}\nonumber\\
&&=(\pi\nu_{S}\Delta/2)^{2}H_{0}^{(1)}(t''\Delta)H_{0}^{(2)}(t'\Delta
)\,,
\end{eqnarray}
where $H_{0}^{(1)}$ and $H_{0}^{(2)}$ are Hankel functions of the
first and second kind, and $\nu_{S}$ is the energy DOS per spin in
the SC at $\mu_{S}$. For times $t',t''>1/\Delta$, the Hankel
functions are rapidly oscillating, since for large $x$ we have
${\rm H_{0}\,^{(1/2)}}(x)\sim\sqrt{2/\pi
x}\,\exp(\pm(ix-(\pi/4)))$. In contrast, the time-dependent phase
in (\ref{current2}) containing the bias $\mu$ suppresses the
integrand in (\ref{current2}) only for times
$|2t-t'-t''|>1/\mu>1/\Delta$. Being  interested only in the
leading order in $\mu/\Delta$, we can assume that $|t|>t',t''$ in
the current formula (\ref{current2}), since the LL correlation
functions are slowly decaying in time with the main contribution
(in the integral) coming from large times $|t|$. In addition,
since $1/\mu
> \Lambda/v_{F}$, we can neglect the term containing the Fermi
velocity $v_{F}$ in (\ref{Greensfunction}). To test the validity
of our approximations we first consider the non-interacting limit
with $K_{\nu}=1$ and $u_{\nu}=v_{F}$, for which an analytic
expression is also available for higher order terms in
$\mu/\Delta$. \\

\section {Non-interacting limit for current $I_{1}$}
\label{nonint} Let us first consider a 1D-Fermi liquid (i.e.
$K_{\nu}=1$ and $u_{\nu}=v_{F}$), and evaluate the integral over
$t$ in Eq. (\ref{current2}) in the non-interacting limit. The
LL-correlation functions simplify to
$G^{1}(t)=(1/2\pi)\lim_{\alpha\to 0}\,1/(\alpha+iv_{F}t)$, and we
are left with the integral
\begin{eqnarray}
&&\int\limits_{-\infty}^{\infty}dt\,e^{i2t\mu}\left\{G^{1}(t-t'')G^{1}(t-t')+G^{1}(t-t'-t'')G^{1}(t)\right\}\nonumber\\
&&=\left(\frac{1}{2\pi}\right)^{2}\int\limits_{-\infty}^{\infty}dt\,e^{i2t\mu}\left\{\frac{1}{\left(\alpha+iv_{F}(t-t'')\right)\left(\alpha+iv_{F}(t-t')\right)}\right.\nonumber\\
&&\left.+\frac{1}{\left(\alpha+iv_{F}(t-t'-t'')\right)\left(\alpha+iv_{F}t\right)}\right\}\Big|_{\alpha\to
0}
\end{eqnarray}
which can be evaluated by closing the integration contour in the
upper complex plane. Inserting then the result into Eq.
(\ref{current2}) we get
\begin{eqnarray}
&&I_{1}=(32e|t_{0}|^{4}/V^{2})\,\lim_{\eta\to
0}\int\limits_{0}^{\infty}dt'\int\limits_{0}^{\infty}dt''\,e^{-\eta(t'+t'')}\nonumber\\
&&\times\,\sum\limits_{{\bf k}{{\bf k}'}}\,u_{\bf k}v_{\bf
k}u_{{\bf k}'}v_{{\bf k}'}\,e^{-iE_{\bf k}t'}\,e^{iE_{{\bf
k}'}t''}e^{i({\bf k}+{\bf k}')\delta{\bf r}}\nonumber\\
&&\times\frac{1}{\pi
v_{F}^{2}}\left\{\frac{\sin\left((t''-t')\mu\right)}{t''-t'}+\frac{\sin\left((t'+t'')\mu\right)}{t'+t''}\right\}.
\label{nonint2}
\end{eqnarray}
The sine-functions in (\ref{nonint2}) can be expanded in powers of
$\mu$, and for $\mu<\Delta$ it is  sufficient to keep just the
leading order term in $\mu$ since the integrals over $t',t''$ have
the form

\begin{equation}
\int_{0}^{\infty}dt\,e^{-(\eta \pm iE_{\bf
k})t}\,t^{n}=\frac{n!}{(\eta\pm iE_{\bf k})^{n+1}}\,, \label{int}
\end{equation}
where $n=1,2,3\cdots$. Since $E_{\bf k}\ge\Delta$, higher powers
in $t',t''$ produce higher powers in $\mu/\Delta$, and, as
expected, we can therefore ignore the dependence on $t',t''$ in
the LL-correlation functions. In contrast to this, when we
consider the current for tunneling of two electrons into the same
(interacting) LL-lead, we will see that the two-particle
correlation function will not allow for such a simplification. In
leading order in $\mu/\Delta$, the integrals over $t',t''$ are
evaluated according to (\ref{int}) with $n=0$, and we get an
(effective) momentum-sum for the SC-correlations $(\sum_{{\bf k}}
\frac{u_{\bf k}v_{\bf k}}{E_{\bf k}} \cos{({\bf k}\cdot\delta {\bf
r})})^{2}$. To evaluate this sum  we use $u_{\bf k}v_{\bf
k}=\Delta/(2 E_{\bf k})$ and linearize the spectrum around the
Fermi-level $\mu_{S}$, since the Fermi energy in the SC
$\varepsilon_{F}>\Delta$, with Fermi wavevector $k_{F}$. We then
obtain ($\delta{r}$ denotes $|\delta {\bf  r}|$)

\begin{equation}
\label{Andreev2} \sum\limits_{{\bf k}} \frac{u_{\bf k}v_{\bf
k}}{E_{\bf k}} \cos{({\bf k}\cdot\delta {\bf r})} =
\frac{\pi}{2}\nu_{S} \frac{\sin(k_{F}\delta r)}{k_{F}\delta r}
e^{-(\delta r/\pi\xi)}\, .
\end{equation}
In (\ref{Andreev2})  we have introduced the coherence length of a
Cooper pair in the SC,  $\xi=v_{F}/\pi \Delta$. We finally obtain
$I_{1}^{0}$, the current $I_{1}$ in the non-interacting limit,

\begin{equation}
I_{1}^{0}=e\pi\gamma^{2}\mu\,\left[\frac{\sin(k_{F}\delta
r)}{k_{F}\delta r}\right]^2 \exp\left(-\frac{2\delta
r}{\pi\xi}\right). \label{I0}
\end{equation}

Here  we have defined $\gamma=4\pi\nu_{S}\nu_{l}|t_{0}|^2/LV$,
which is the dimensionless conductance per spin to tunnel from the
SC to the LL-leads. The non-interacting DOS of the LL per spin
$\nu_{l}$ is given by $\nu_{l}=L/\pi v_{F}$. [We remark in passing
that this result agrees with a $T$-matrix calculation in the
energy domain \cite{unpublished}. In this case we sum explicitly
over the final states, given by a singlet
$|f\rangle=(1/\sqrt{2})[a_{1{ p}\uparrow}^{\dagger}a_{2 {
q}\downarrow}^{\dagger}- a_{1{ p}\downarrow}^{\dagger}a_{2{
q}\uparrow}^{\dagger}]|i\rangle$, where the $a$-operators describe
electrons in  a non-interacting 1D Fermi-liquid.  Note that
triplet states  are excluded as final states since our Hamiltonian
$H$ does not change the total spin.] We see that the current
$I_{1}^{0}$ gets exponentially suppressed on the scale of $\xi$,
if the tunneling of the two (coherent) electrons takes place from
different points ${\bf r}_{1}$ and ${\bf r}_{2}$ of the SC. For
conventional s-wave SC the coherence length $\xi$ is typically on
the order of micrometers and therefore this poses not severe
experimental restrictions. Thus, in the  regime of interest
$\delta r<\xi$, the suppression of the current $I_{1}^{0}$ is only
polynomial, i.e. $\propto (1/k_{F}\delta r)^{2}$. It was shown
\cite{VanWees} that a superconductor on top of a two-dimensional
electron gas (2DEG) can induce superconductivity (by the proximity
effect) in the 2DEG with a finite order parameter. The 2DEG then
becomes an effective two-dimensional (2D) SC. More recently, it
was suggested that superconductivity should also be present in
ropes of single-walled carbon nanotubes \cite{Bouchiat}, which are
one-dimensional (1D) systems. It is therefore interesting to
calculate (\ref{Andreev2}) also in 2D and 1D. In the case of a 2D
SC  we evaluate $\sum_{{\bf k}} \frac{u_{\bf k}v_{\bf k}}{E_{\bf
k}} \cos{({\bf k}\cdot\delta {\bf r})}$  in leading order in
$\delta r/\pi\xi$, and we find
\begin{eqnarray}
&&\sum\limits_{{\bf k}({\tiny 2D})} \frac{u_{\bf k}v_{\bf
k}}{E_{\bf k}}\cos{({\bf k}\cdot\delta {\bf r})}\nonumber\\
&&=\frac{\pi}{2}\nu_{S}\left(J_{0}(k_{F}\delta
r)+2\sum\limits_{\nu=1}^{\infty}\,\frac{J_{2\nu}(k_{F}\delta
r)}{\pi\nu}\right), \label{Andreev3}
\end{eqnarray}
where $J_{\nu}(x)$ denotes the Bessel function of order $\nu$. For
large $k_{F}\delta r$, we get
 $J_{\nu}(k_{F}\delta r)\sim\sqrt{2/\pi k_{F}\delta r}\,\cos(k_{F}\delta
 r-(\nu\pi/2)-(\pi/4))$,
 which allows an approximation of the right-hand side of ({\ref{Andreev3})
 for large $k_{F}\delta r$ by $(\pi/2)\nu_{S}\sqrt{2/\pi k_{F}\delta r}\cos(k_{F}\delta r-(\pi/4))(1-(2/\pi)\ln 2)$.
 This result is exact to leading order in an expansion in $1/k_{F}\delta r$.  So
 asymptotically, the current decays only $\propto 1/k_{F}\delta r$. For $\delta r=0$,
 the bracket on the right-hand side of ({\ref{Andreev3}) becomes 1 as in the 3D-case.

In the case of a 1D-SC we obtain
\begin{equation}
\sum\limits_{{\bf k}(1D)} \frac{u_{\bf k}v_{\bf k}}{E_{\bf
k}}\cos{({\bf k}\cdot\delta {\bf r})}=
\frac{\pi}{2}\nu_{S}\cos(k_{F}\delta r)\,e^{-(\delta r/\pi\xi)},
\end{equation}
where there are only oscillations and no decay of the Andreev
amplitude (for $\delta r/\pi \xi<1)$. We see that the suppression
of the current due to a finite separation of the tunneling points
on the SC can be reduced considerably (or even excluded
completely) by going over to lower dimensional SCs.\\

\section{ current $I_{1}$ including interaction. }

We now are ready to treat the interacting case. Having obtained
confidence in our approximation schemes from the non-interacting
case above, we can neglect now the $t',t''$ dependence of the
LL-correlation function appearing in (\ref{current2}), valid in
leading order in $\mu/\Delta$. In this limit the $t$-integral
considerably simplifies to
\begin{eqnarray}
&&(2\pi)^{2}\int_{-\infty}^{\infty}dt\,e^{i(2t-t'-t'')\mu}\nonumber\\
&&\times\left\{G^{1}(t-t'')G^{1}(t-t')+G^{1}(t-t'-t'')G^{1}(t)\right\}\nonumber\\
&&\nonumber\\ &&\sim
\frac{2\Lambda^{2(\gamma_{\rho}+\gamma_{\sigma})}}{\prod_{\nu=\rho,\sigma}u_{\nu}^{2\gamma_{\nu}+1}}\,\int_{-\infty}^{\infty}dt\,\frac{e^{i2\mu
t}}{\prod_{\nu=\rho,\sigma}\left((\Lambda/u_{\nu})+it\right)^{2\gamma_{\nu}+1}}.\nonumber\\
&&
\label{Int1}
\end{eqnarray}
An analytical expression for this integral is available
\cite{G/R}, and given in Appendix B. The treatment of the
remaining integrals over $t',t''$ and the calculation of the
Andreev contribution is the same as in the non-interacting case
and
 we obtain for the current $I_{1}$, in leading order in $\mu/\Delta$ and in the small parameters $2\Lambda\mu/u_{\nu}$,
\begin{equation}
I_{1}=\frac{I_{1}^{0}}{\Gamma(2\gamma_{\rho}+2)}\frac{v_{F}}{u_{\rho}}\left[\frac{2\mu\Lambda}{u_{\rho}}\right]^{2\gamma_{\rho}}.
\label{I_{1}}
\end{equation}
In (\ref{I_{1}}) we used $K_{\sigma}=1$ and $u_{\sigma}=v_{F}$.
The interaction suppresses the current considerably and the bias
dependence has its characteristic non-linear form, $I_{1}\propto
(\mu)^{2\gamma_{\rho}+1}$, with an interaction dependent exponent
$2\gamma_{\rho}+1$. The parameter $\gamma_{\rho}$ is the exponent
for tunneling into the bulk of a single LL, i.e.
$\rho(\varepsilon)\sim|\varepsilon|^{\gamma_{\rho}}$
\cite{Schulz}, where $\rho(\varepsilon)$ is the single particle
DOS. Note that the current $I_{1}$ does not show a dependence on
the  correlation time  $1/\Delta$, which is a measure for the time
separation between the two electron tunneling-events. This is so
since the two partners of a Cooper pair tunnel to {\em different}
LL leads with no interaction-induced correlations between the
leads.
\section{ Current $I_{2}$ for tunneling of two electrons into the same lead }
The main new feature for the case where two electrons, originating
from an Andreev process, tunnel into the same lead, is now that
the 4-point correlation function of the LL  no longer factorizes
as was the case before when the two electrons tunnel into
different leads (see (\ref{totcorrelation})). In addition the two
electrons will tunnel into the lead prefarably from the same
spatial point on the SC, i.e. $\delta r=0$. We denote by $I_{2}$
the current for coherent transport of two electrons into the {\em
same} lead, either lead 1 {\em or} lead 2. It can be written in a
similar way as $I_{1}$ (see (\ref{current1})) with the difference
that now  we consider final states with two additional electrons
(of opposite spin) in the same lead (either 1 or 2) compared to
the initial state. We then obtain for $I_{2}$,

\begin{eqnarray}
&&I_{2}=4e\lim_{\eta \to
0}\sum\limits_{s,s'}\,\int\limits_{-\infty}^{\infty}dt\int\limits_{0}^{\infty}dt'\int\limits_{0}^{\infty}dt''\,e^{-\eta(t'+t'')}\nonumber\\
&&e^{i(2t-t'-t'')\mu}\,\langle
H_{Tn-s'}^{\dagger}(t-t'')H_{Tns'}^{\dagger}(t)H_{Tn-s}(t')H_{Tns}\rangle,\nonumber\\
&& \label{current2f}
\end{eqnarray}
where we have used that the leads 1 and 2 are identical which
results in an additional  factor of two. Again, the thermal
average is to be taken at $T=0$, and this groundstate expectation
value factorizes into  a SC-part times a LL-part. However, the
LL-part does not factorize anymore due to strong correlations
between the two tunneling electrons. We obtain in this case
\begin{eqnarray}
&&\sum\limits_{s,s'}\,\langle
H_{Tn-s'}^{\dagger}(t-t'')H_{Tns'}^{\dagger}(t)H_{Tn-s}(t')H_{Tns}\rangle\nonumber\\
&&=|t_{0}|^{4}\,\sum\limits_{s}\,\langle\psi_{ns}(t-t'')\psi_{n-s}(t)\psi_{n-s}^{\dagger}(t')\psi_{ns}^{\dagger}\rangle\nonumber\\
&&\times\,\langle\Psi_{s}^{\dagger}({\bf
r}_{n},t-t'')\Psi_{-s}^{\dagger}({\bf r}_{n},t)\Psi_{-s}({\bf
r}_{n},t')\Psi_{s}({\bf r}_{n})\rangle\nonumber\\ &&\nonumber\\
&&+|t_{0}|^{4}\,\sum\limits_{s}\,\langle\psi_{n-s}(t-t'')\psi_{ns}(t)\psi_{n-s}^{\dagger}(t')\psi_{ns}^{\dagger}\rangle\nonumber\\
&&\times\,\langle\Psi_{-s}^{\dagger}({\bf
r}_{n},t-t'')\Psi_{s}^{\dagger}({\bf r}_{n},t)\Psi_{-s}({\bf
r}_{n},t')\Psi_{s}({\bf r}_{n})\rangle. \label{correlation2}
\end{eqnarray}
The 4-point correlation functions  for the SC in
(\ref{correlation2}) are the same as in (\ref{totcorrelation}) for
the case when the two electrons tunnel into different leads,
except that now $\delta {\bf r}=0$. The (normalized) 4-point
correlation functions in (\ref{correlation2}) for the LL are
$G_{rr'1}^{2}(t,t',t'')\equiv\langle
\psi_{nrs}(t-t'')\psi_{nr'-s}(t)\psi_{nr'-s}^{\dagger}(t')\psi_{nrs}^{\dagger}\rangle/\langle
\psi_{nrs}(t-t'')\psi_{nrs}^{\dagger}\rangle\\\langle\psi_{nr'-s}(t-t')\psi_{nr'-s}^{\dagger}\rangle$,
which can be calculated using similar methods as described above
for the single-particle correlation function. After some
calculation we get
\begin{eqnarray}
&&\,\,\,G_{rr'1}^{2}(t,t',t'')=\nonumber\\ &&\nonumber\\
&&\prod_{\nu=\rho,\sigma}\left(\frac{\Lambda-iu_{\nu}t''}{\Lambda+iu_{\nu}(t-t'-t'')}\right)^{\gamma_{\nu
rr'}}\left(\frac{\Lambda+iu_{\nu}t'}{\Lambda+iu_{\nu}t}\right)^{\gamma_{\nu
rr'}}\nonumber\\
&&\times\,\left[\frac{(\Lambda+iu_{\sigma}(t-t'-t''))(\Lambda+iu_{\sigma}t)}{(\Lambda+iu_{\rho}(t-t'-t''))(\Lambda+iu_{\rho}t)}\right]^{\frac{1+rr'}{4}}\nonumber\\
&&\times\left[\frac{(\Lambda-iu_{\rho}t'')(\Lambda+iu_{\rho}t')}{(\Lambda-iu_{\sigma}t'')(\Lambda+iu_{\sigma}t')}\right]^{\frac{1+rr'}{4}}
\label{4PunktLL1}
\end{eqnarray}
where $\gamma_{\nu
rr'}=\xi_{\nu}((1/K_{\nu})+rr'K_{\nu}-(1+rr'))/4$ with
$\xi_{\rho/\sigma}=\pm 1$. The exponent $\gamma_{\nu rr'}$ is
related to $\gamma_{\nu}$, introduced in the single-particle
correlation function (\ref{Greensfunction}) via $\gamma_{\nu
rr'}=\xi_{\nu}\,\gamma_{\nu}$ for $r=r'$, and $\gamma_{\nu
rr'}=\xi_{\nu}\,(2\gamma_{\nu}+1-K_{\nu})/2$ for $r\neq r'$. For
the other sequence $G_{rr'2}^{2}=\langle
\psi_{nr-s}(t-t'')\psi_{nr's}(t)\psi_{nr-s}^{\dagger}(t')\psi_{nr's}^{\dagger}\rangle/\langle
\psi_{nr-s}(t-t'-t'')\psi_{nr-s}^{\dagger}\rangle\langle\psi_{nr's}(t)\psi_{nr's}^{\dagger}\rangle$,
we obtain
\begin{eqnarray}
&&\,\,\,G_{rr'2}^{2}(t,t',t'')=\nonumber\\ &&\nonumber\\
&&-\prod_{\nu=\rho,\sigma}\,\left(\frac{\Lambda-iu_{\nu}t''}{\Lambda+iu_{\nu}(t-t'')}\right)^{\gamma_{\nu
rr'}}\left(\frac{\Lambda+iu_{\nu}t'}{\Lambda+iu_{\nu}(t-t')}\right)^{\gamma_{\nu
rr'}}\nonumber\\
&&\times\left[\frac{(\Lambda+iu_{\sigma}(t-t''))(\Lambda+iu_{\sigma}(t-t'))}{(\Lambda+iu_{\rho}(t-t''))(\Lambda+iu_{\rho}(t-t'))}\right]^{\frac{1+rr'}{4}}\nonumber\\
&&\times\left[\frac{(\Lambda-iu_{\rho}t'')(\Lambda+iu_{\rho}t')}{(\Lambda-iu_{\sigma}t'')(\Lambda+iu_{\sigma}t')}\right]^{\frac{1+rr'}{4}}.
\label{4PunktLL2}
\end{eqnarray}
We remark that contributions from other combinations of left- and
right- movers, as indicated in (\ref{4PunktLL1}) and
(\ref{4PunktLL2}), are negligible. A contribution like
$\langle\psi_{nr's}(t-t'')\psi_{nr-s}(t)\psi_{nr'-s}^{\dagger}(t')\psi_{nrs}^{\dagger}\rangle_{r\neq
r'}$ is only non-zero if spin exchange between right- and
left-movers is possible, but this is a backscattering process
which we explicitly exclude. Using (\ref{current2f}) together with
(\ref{4PunktLL2}), we obtain a formal expression for $I_{2}$ (with
$\delta r=0$)
\begin{widetext}
\begin{eqnarray}
I_{2}&=&4e\left(\frac{\pi\nu_{S}\Delta\,|t_{0}|^{2}}{V}\right)^{2}\,\lim_{\eta \to 0}\int\limits_{-\infty}^{\infty}dt\int\limits_{0}^{\infty}dt'\int\limits_{0}^{\infty}dt''\,e^{-\eta(t'+t'')}\,e^{i(2t-t'-t'')\mu}\,{\rm H_{0}^{(1)}}(t''\Delta){\rm H_{0}^{(2)}}(t'\Delta)\nonumber\\
&&\nonumber\\
&\times&\sum\limits_{b=\pm 1}\,\big(G_{b1}^{2}(t,t',t'')\,G^{1}(t-t'')\,G^{1}(t-t')-\,G_{b2}^{2}(t,t',t'')\,G^{1}(t-t'-t'')\,G^{1}(t)\big).
\label{currentI22}
\end{eqnarray}
\end{widetext}
In (\ref{currentI22}) the meaning of the summation index is
$b\equiv +1$ for $rr'=+1$, and  $b\equiv -1$ for $rr'=-1$. We
proceed to evaluate the current $I_{2}$ with $K_{\sigma}=1$ and
$u_{\sigma}=v_{F}$, i.e. $\gamma_{\sigma rr'}=0$. Since
$\gamma_{\rho rr'}>0$, we see from the first line in
(\ref{4PunktLL1}) and (\ref{4PunktLL2}) that for
$|t|>\Lambda/u_{\rho},t',t''$, the full 4-point correlation
function
 is suppressed by a factor $\sim [t't''/(t^{2})]^{\gamma_{\rho rr'}}$ compared to its factorization approximation.
To calculate the current $I_{2}$ we assume that the time scales
$\Lambda/u_{\rho}$ and $\Lambda/v_{F}$ are the smallest ones in
the problem. The times $\Lambda/u_{\rho}$ and $\Lambda/v_{F}$ are
both on the order of the inverse Fermi energy in the LL, which is
larger than the energy gap $\Delta$ and the bias $\mu$. By
applying the same arguments as in Section \ref{current1S} for the
current $I_{1}$, we approximate the current $I_{2}$, assuming
$|t|> t',t''>\Lambda/v_{F},\Lambda/u_{\rho}$, which is accurate in
leading order in the small parameters $\mu/\Delta$,
$\Lambda\Delta/u_{\nu}$ and $\Lambda\mu/u_{\nu}$. In this limit we
obtain for the two-particle correlation functions
$G_{rr'1}^{2}=-G_{rr'2}^{2}=u_{\rho}^{2\gamma_{\rho
rr'}}\,\,(t't'')^{\gamma_{\rho
rr'}}/(\Lambda+iu_{\rho}t)^{2\gamma_{\rho rr'}}$. The current
$I_{2}$ for tunneling of two electrons into the {\em same} lead
$1$ or $2$ then becomes (for $\delta r=0$)
\begin{widetext}
\begin{eqnarray}
I_{2}&=&2e\left(\frac{2\pi \nu_{S} \Delta |t_{0}|^{2}}{V}\right)^{2}\frac{\Lambda^{2\gamma_{\rho}}}{u_{\rho}^{2\gamma_{\rho}+1}v_{F}}\,\lim_{\eta\to 0}\,\sum\limits_{b=\pm 1}\,\int\limits_{0}^{\infty}dt'\int\limits_{0}^{\infty}dt''e^{-\eta(t'+t'')}\,\left(t't''\right)^{\gamma_{\rho b}}{\rm H_{0}^{(1)}}(t''\Delta){\rm H_{0}^{(2)}}(t'\Delta)\nonumber\\
&\times&\frac{1}{(2\pi)^{2}}\,\int\limits_{-\infty}^{\infty}dt\,\frac{e^{i2\mu t}}{\left(\frac{\Lambda}{u_{\rho}}+it\right)^{2\gamma_{\rho b}+2\gamma_{\rho}+1}\left(\frac{\Lambda}{v_{F}}+it\right)}\,.
\label{currentI2}
\end{eqnarray}
\end{widetext}
Again, the integrals appearing in (\ref{currentI2}) can be
evaluated analytically \cite{G/R} with the results given in
Appendix B. Note that according to the two-particle LL-correlation
functions (\ref{4PunktLL1}) and (\ref{4PunktLL2}), we find that
the dynamics coming from the delay times $t'$ and $t''$ cannot be
neglected anymore, as was done in \cite{Fazio}.  We evaluate
(\ref{currentI2}) in leading order in $2\mu\Lambda/u_{\nu}$ and
finally obtain for the current $I_{2}$
\begin{equation}
I_{2}=I_{1}\sum\limits_{b=\pm
1}\,A_{b}\,\left(\frac{2\mu}{\Delta}\right)^{2\gamma_{\rho b}}.
\label{currentI222}
\end{equation}

The interaction dependent constant $A_{b}$ in (\ref{currentI222}) is
given by
\begin{equation}
A_{b}=\frac{2^{2\gamma_{\rho
b}-1}}{\pi^{2}}\frac{\Gamma(2\gamma_{\rho}+2)}{\Gamma(2\gamma_{\rho
b}+2\gamma_{\rho}+2)}\,\Gamma^{4}\left(\frac{\gamma_{\rho
b}+1}{2}\right),
\end{equation}
which is decreasing for increasing the interactions in the LL leads and $\Gamma(x)$ is the Gamma function. We remark that in
(\ref{currentI222}) the current $I_{1}$ is to be taken at $\delta
r=0$. The non-interacting limit, $I_{2}=I_{1}=I_{1}^{0}$, is
recovered by putting $\gamma_{\rho}=\gamma_{\rho b}=0$, and
$u_{\rho}=v_{F}$. The result for $I_{2}$ shows that the unwanted
injection of two electrons into the same lead is suppressed
compared to $I_{1}$ by a factor of $A_{+}(2\mu/\Delta)^{2\gamma_{\rho
+}}$ if both electrons are injected into the same branch (left or
right movers), or by $A_{-}(2\mu/\Delta)^{2\gamma_{\rho -}}$ if the two
electrons travel in different directions.  Since it holds that
$\gamma_{\rho -}=\gamma_{\rho +}+(1-K_{\rho})/2>\gamma_{\rho +}$,
it is more favorable that two electrons travel in the same
direction than in opposite directions. The suppression of the
current $I_{2}$ by $1/\Delta$ shows very nicely the two-particle
correlation effect for the coherent tunneling of two electrons
into the same lead. The larger $\Delta$, the shorter is the delay
time between the arrivals of the two partner electrons of a given
Cooper pair, and, in turn, the more the second electron will be
influenced by the presence of the first one already  in the LL. By
increasing the bias $\mu$ the electrons can tunnel faster through
the barrier due to more channels becoming available into which the
electron can tunnel, and therefore the effect of $\Delta$ is less
pronounced. Also note that this correlation effect disappears when
interactions  are absent  in the LL  ($\gamma_{\rho}=\gamma_{\rho
b}=0$).

\section{Efficiency and Discussion}
We have established now that there exists indeed the suppression
for tunneling of two spin-entangled electrons into the same
LL-lead compared to the desired process where the two electrons
tunnel into different leads. However, we have to take into account
that the process into different leads suffers also a suppression
due to a finite tunneling separation $\delta r$ of the two
electrons forming a Cooper pair in the SC. In  Section
\ref{nonint}  we showed  that this suppression can be considerably
reduced if one uses effectively low-dimensional SCs. To estimate
the efficiency of the entangler we form the ratio $I_{1}/I_{2}$
and demand that it is larger than one. This requirement is
fulfilled if approximately
\begin{equation}
A_{+}\left(\frac{2\mu}{\Delta}\right)^{2\gamma_{\rho +}}\,<\,1/(k_{F}\delta r)^{d-1}\,,
\end{equation}
 where $d$ is the dimension of the SC, and it is assumed that the
 coherence length $\xi$ of the SC is large compared to $\delta r$.
 The leading term of $I_{2}$ is proportional to $(2\mu/\Delta)^{2\gamma_{\rho +}}$
 describing the power-law suppression, with exponent $2\gamma_{\rho +}=2\gamma_{\rho}$,
 of the process where two electrons, entering the same lead, will propagate in the same
 direction. The exponent $\gamma_{\rho +}$ is the exponent for the single-particle
 tunneling-DOS from a metal (SC) into the center (bulk) of a LL.
 Experimentally
 accessible systems   which exhibit LL-behavior are metallic carbon nanotubes.
 It was pointed out \cite{Egger,Kane} that the long range part of the Coulomb
 interaction, which is dominated by  forward scattering
events with small momentum transfer, can lead to LL behavior in
carbon nanotubes with very small values of $K_{\rho}\sim 0.2-0.3$,
as measured experimentally \cite{Bockrath,Nygard} and predicted
theoretically \cite{Kane}. This would correspond to an exponent
$2\gamma_{\rho}\sim 0.8-1.6$, which seems very promising. In
addition, single-wall nanotubes show similar tunneling exponents
as derived here.
 The tunneling DOS for a single-wall nanotube is predicted to
 be $\rho(\varepsilon)\sim |\varepsilon|^{\eta}$
 with $\eta=(K_{\rho}^{-1}+K_{\rho}-2)/8$ \cite{Egger,Kane}, which is half of $\gamma_{\rho}$,
 and was measured \cite{Bockrath,Nygard} to be $\sim 0.3-0.4$. Similar values
 were found  also in  multiwall-nanotubes \cite{Schoenenberger}. It is known
 that the power-law suppression of the single-particle DOS is even larger if
 one considers tunneling into the {\em end} of a LL. For single wall nanotubes
 one finds $\eta_{end}=(K_{\rho}^{-1}-1)/4> \eta$ \cite{Egger,Kane}, or for
 conventional LL-theory again an enhancement by a factor of two \cite{Balents}.
 We therefore expect to get an even stronger suppression if the Cooper pairs tunnel
 into the end of the LLs.
We remark that the non-locality of the two electrons could be
probed via the Aharonov-Bohm oscillations in the current, when the
leads 1,2 are formed into a loop enclosing a magnetic flux. Due to
the different paths which the electrons can choose to go around
the loop, we expect to see $h/e$ and $h/2e$ oscillation periods,
as a function of magnetic flux, in the current like for
non-interacting leads \cite{RSL}. Interference of contributions
where the two electrons travel through different leads with
contributions where they travel through the same lead then lead to
the $h/e$ oscillations, whereas interference of contributions
where both electrons travel through the same arm 1 or 2 of the
loop lead to the $h/2e$ oscillations. The amplitudes of these
oscillations must be related to the currents describing the
interfering processes. We expect that the $h/e$ oscillation
contribution should be $\propto (I_{1}I_{2})^{\alpha}$ and the
$h/2e$ oscillation contribution should be $\propto
I_{2}^{2\alpha}$, with an exponent $\alpha$ that has to be
determined by explicit calculations. In the non-interacting limit
$\alpha$ should be $1/2$ \cite{RSL}. The different periods then
allow for an experimental test of how successful the separation of
the two electrons is. For instance, if the two electrons only can
tunnel into the same lead, e.g. if $k_{F}\delta r$ is too large or
the interaction in the leads too weak, then $I_{1}\sim 0$ and we
would only see the $h/2e$ oscillations in the current. The
determination of the precise value of the exponents will be
deferred to another publication since it requires a separate
calculation including finite size properties of the LL along the
lines discussed in \cite{GeLo}.

\section{Decay of the electron-singlet due to LL-interactions}

We have shown in the preceding sections that the interaction in a
LL-lead can help to separate two spin-entangled electrons so that
the two electrons enter {\em different} leads. A natural
question  then arises: what is the lifetime of a (non-local)
spin-singlet state formed of two electrons  which are injected
into different LL-leads, one electron per lead? To address this
issue we introduce the following  correlation function
\begin{equation}
P({\bf r},t)=\left|\langle S({\bf r},t)|S(0,0)\rangle\right|^{2}.
\end{equation}
This function is the probability density that a singlet state,
injected at point ${\bf r}\equiv (x_{1},x_{2})=0$ and at time
$t=0$, is found at some later time $t$ and at point ${\bf r}$.
Therefore, $P({\bf r},t)$ is a measure of how much of the initial
singlet state remains after the two injected electrons have
interacted with all the other electrons in the LL during the time
interval $t$. Here
\begin{eqnarray}
|S({\bf
r},t)\rangle&=&\sqrt{\pi\alpha}\,[\psi_{1\uparrow}^{\dagger}(x_{1},t)\psi_{2\downarrow}^{\dagger}(x_{2},t)\nonumber\\
&-&\psi_{1\downarrow}^{\dagger}(x_{1},t)\psi_{2\uparrow}^{\dagger}(x_{2},t)]|0\rangle
\label{singlet}
\end{eqnarray}
is the electron singlet state created on top of the LL
groundstates. The extra normalization factor $\sqrt{2\pi\alpha}$
is introduced to guarantee $\int d{\bf r}\,P({\bf r},t)=1$ in the
non-interacting limit and corresponds to the replacement of
$\psi_{ns}$ by $(2\pi\alpha)^{1/4}\psi_{ns}$. The singlet-singlet
correlation function factorizes into two single-particle Green's
functions due to negligible interaction between the leads $1$ and
$2$. Therefore we have $P({\bf
r},t)=(2\pi\alpha)^{2}\prod_{n}\left|\langle\psi_{ns}(x_{n},t)\psi_{ns}^{\dagger}(0,0)\rangle\right|^{2}$,
with
$\langle\psi_{ns}(x_{n},t)\psi_{ns}^{\dagger}(0,0)\rangle=\sum_{r=\pm}\,e^{ik_{F}rx_{n}}G^{1}_{nrs}(x_{n},t)$.
For simplicity we just study the slow spatial variations of
$\left|\langle\psi_{ns}(x_{n},t)\psi_{ns}^{\dagger}(0,0)\rangle\right|^{2}$
and obtain with (\ref{Greensfunction})
\begin{eqnarray}
&&(2\pi)^{2}\,\left|\langle\psi_{ns}(x_{n},t)\psi_{ns}^{+}(0,0)\rangle\right|^{2}\nonumber\\
&&\nonumber\\ &&=\sum\limits_{r=\pm}\lim\limits_{\alpha\to
0}\frac{\Lambda^{2}+(v_{F}t-rx_{n})^{2}}{\alpha^{2}+(v_{F}t-rx_{n})^{2}}\prod_{\nu=\rho,\sigma}\frac{1}{\sqrt{\Lambda^{2}+(rx_{n}-u_{\nu}t)^{2}}}\nonumber\\
&&\times\left(\frac{\Lambda^{4}}{\left(\Lambda^{2}+x_{n}^{2}+(u_{\nu}t)^{2}\right)^{2}+\left(2u_{\nu}t\Lambda\right)^{2}}\right)^{\gamma_{\nu}/2}.
\label{fidelity}
\end{eqnarray}

If we use $\pi\delta(x)=\lim_{\alpha\to
0}\alpha/(\alpha^{2}+x^{2})$ we can then write the remaining
probability of the singlet as
\begin{equation}
P({\bf
r},t)=\prod_{n}\frac{1}{2}\sum\limits_{r=\pm}F(t)\,\delta(x_{n}-rv_{F}t)
\label{P}
\end{equation}
with  a time decaying weight factor of the $\delta$-function
\begin{eqnarray}
F(t)&=&\prod_{\nu=\rho,\sigma}\sqrt{\frac{\Lambda^{2}}{\Lambda^{2}+(v_{F}-u_{\nu})^{2}t^{2}}}\nonumber\\
&\times&\left(\frac{\Lambda^{4}}{\left(\Lambda^{2}+(v_{F}t)^{2}+(u_{\nu}t)^{2}\right)^{2}+\left(2\Lambda
u_{\nu}t\right)^{2}}\right)^{\gamma_{\nu}/2}.
\end{eqnarray}

Without interaction we have $F(t)=1$, which means that there is no
decay of the singlet state. As interactions  are turned on, we see
that for times $t>\Lambda/u_{\nu}$ the singlet state decays in
time with  approximately
$F(t)\sim\prod_{\nu=\rho,\sigma}\frac{\Lambda}{\sqrt{\Lambda^{2}+(u_{\nu}-v_{F})^{2}t^{2}}}
\left(\frac{\Lambda}{t\sqrt{v_{F}^{2}+u_{\nu}^{2}}}\right)^{2\gamma_{\nu}}$.
This result together with (\ref{P})  shows that charge and spin of
an electron propagate with velocity $v_{F}$, whereas charge (spin)
excitations of the LL propagate with $u_{\rho}$ ($u_{\sigma}$). In
addition, we see that the probability $P({\bf r},t)$ shows an
additional power-law decay
$\sim(\frac{\Lambda}{t\sqrt{v_{F}^{2}+u_{\nu}^{2}}})^{2\gamma_{\nu}}$,
with an interaction dependent exponent. We will show in the next
section that although the singlet gets destroyed due to
interactions, we still can observe charge and spin of the initial
singlet via the spin and charge density fluctuations of the LL.

\section{Propagation of charge and spin}
The charge and spin propagation as a function of time in a state
$|\Psi\rangle$ can be described by the correlation function
$\langle\Psi|\rho(x,t)|\Psi\rangle$ for the charge, and
$\langle\Psi|\sigma_{z}(x,t)|\Psi\rangle$ for the spin. The
normal-ordered charge density operator for LL n is
$\rho_{n}(x_{n})=\sum_{s}\,:\psi_{ns}^{\dagger}(x_{n})\psi_{ns}(x_{n})
:=\sum_{sr}\,:\psi_{nsr}^{\dagger}(x_{n})\psi_{nsr}(x_{n}):$, if
we only consider the slow spatial variations of the density
operator. Similarly, the normal-ordered spin density operator in
z-direction is
$\sigma_{n}^{z}(x_{n})=\sum_{sr}s\,:\psi_{nsr}^{\dagger}(x_{n})\psi_{nsr}(x_{n}):$\,.
These density fluctuations can be expressed in a bosonic form (see
Appendix A) as
\begin{equation}
\rho_{n}(x_{n})=\frac{\sqrt{2}}{\pi}\,\partial_{x}\phi_{n\rho}(x_{n})\,
\label{charge}
\end{equation}
and for the spin
\begin{equation}
\sigma_{n}^{z}(x_{n})=\frac{\sqrt{2}}{\pi}\,\partial_{x}\phi_{n\sigma}(x_{n}).
\label{spin}
\end{equation}

We now consider a state
$|\Psi\rangle=\psi_{nsr}^{\dagger}(x_{n})|0\rangle$ where we
inject an electron at time $t=0$ into branch $r$ on top of the LL
groundstate in lead $n$ and calculate the time dependent charge
and spin density fluctuations according to $\langle
0|\psi_{nsr}(x_{n})\rho_{n}(x_{n}',t)\psi_{nsr}^{\dagger}(x_{n})|0\rangle$
for the charge and similar for the spin $\langle
0|\psi_{nsr}(x_{n})\sigma_{n}^{z}(x_{n}',t)\psi_{nsr}^{\dagger}(x_{n})|0\rangle$.
If we express the bosonic fieldoperators $\phi_{n\nu}$ and
$\theta_{n\nu}$ in terms of the boson modes shown in (\ref{bogo2})
and (\ref{bogo3}) and the Fermi operators according to the
bosonization dictionary (\ref{bosonization}) we obtain for the
charge fluctuations
\begin{eqnarray}
&&\left\langle
0|\psi_{nsr}(x_{n})\rho_{n}(x_{n}',t)\psi_{nsr}^{\dagger}(x_{n})|0\right\rangle\nonumber\\
&&=\frac{1}{2}(1+rK_{\rho})\delta\left(x_{n}'-x_{n}-u_{\rho}t\right)\nonumber\\
&&+\frac{1}{2}(1-rK_{\rho})\delta\left(x_{n}'
-x_{n}+u_{\rho}t\right)\, ,
 \label{flukt1}
\end{eqnarray}
and for the spin fluctuations
\begin{eqnarray}
&&\left\langle
0|\psi_{nsr}(x_{n})\sigma_{n}^{z}(x_{n}',t)\psi_{nsr}^{\dagger}(x_{n})|0\right\rangle\nonumber\\
&&=\frac{s}{2}(1+rK_{\sigma})\delta\left(x_{n}'-x_{n}-u_{\sigma}t\right)\nonumber\\
&&+\frac{s}{2}(1-rK_{\sigma})\delta\left(x_{n}'-x_{n}+u_{\sigma}t\right).
\label{flukt2}
\end{eqnarray}
The results (\ref{flukt1}) and (\ref{flukt2}) are obtained by sending $\Lambda\rightarrow 0$ and
by using the same normalization convention for the electron
operators as in (\ref{singlet}). We see that in contrast to the
singlet, the charge and spin density fluctuations in the LL
created by the injected electron do not decay and show a pulse
shape with no dispersion in time. This is due to the linear energy
dispersion relation of the LL-model. In carbon nanotubes such a
highly linear dispersion relation is indeed realized, and,
therefore,  nanotubes should be well suited for spin transport.
Another interesting effect that shows up in (\ref{flukt1}) and
(\ref{flukt2}) is the different velocities of spin and charge,
which is known as spin-charge separation.  It would be interesting
to test Bell inequalities \cite{Bell} via spin-spin correlation
measurements between the two LL-leads and see if the initial
entanglement of the spin singlet is still observable in the spin
density-fluctuations. Although detection of single spins with
magnitudes on the order of electron spins has still not been
achieved, magnetic resonance force microscopy (MFRM) seems to be
very promising in doing so \cite{Rugar}. Another scenario is to
use the LL just as an intermediate medium which is needed to first
separate the two electrons of a Cooper pair and then to take them
(in general other electrons) out again into two (spatially
separated) Fermi liquid leads where the (possibly reduced)  spin
entanglement could be measured via the current noise in a
beamsplitter experiment \cite{BLS}. Similarly, to test Bell
inequalities one can make then use of measuring spin via the
charge of the electron \cite{Loss97,Recher,EL}.
 In this context we finally note that
the decay of the singlet state given by (\ref{P}) sets in almost
immediately after the injection into the LLs (the time scale is
approximately the inverse of the Fermi energy), but at least at
zero temperature, the suppression is only polynomial in time,
which suggests that some fraction of the singlet state can still
be recovered.

\section{conclusions}
We proposed an s-wave superconductor (SC), coupled to two
spatially separated Luttinger liquid (LL) leads, as an entangler
for electron spins. We showed that the strong correlations present
in the LL can be used to separate two electrons, forming a
spin-singlet state, which originate from an Andreev tunneling
process of a Cooper pair from the SC to the leads. We have shown
that the coherent tunneling of two electrons into the same lead is
suppressed by a characteristic power law in the small parameter
$\mu/\Delta$, where $\mu$ is the applied bias between the SC and
the LL-leads, and $\Delta$ is the gap in the SC. On the other
hand, when the two electrons tunnel into different leads, the
current is suppressed by the initial separation of the two
electrons. This suppression, however, can be considerably reduced
by going over to effective lower-dimensional SC. We also addressed
the question of how much of the initial singlet can be taken out
of the LL  at some later time, and we found that the probability
is decreasing in time, again with a power-law (at zero
temperature). Nevertheless, the spin information can still be
transported through the wires by means of the (proper) spin
excitations of the LL.

While preparing this manuscript we have learned of related and
independent efforts by S. Vishveshwara et al.\cite{Smitha} who
consider a similar setup as proposed here thereby arriving at
similar conclusions.

\acknowledgments We thank W. Belzig, C. Bruder, F. Meier, and E.V.
Sukhorukov for useful discussions. We also gratefully acknowledge
discussions with S. Vishveshwara, C. Bena, L. Balents, and M.P.A.
Fisher. This work has been supported by the Swiss NSF, DARPA, and
ARO.

\appendix
\section{Finite Size Diagonalization of the LL-Hamiltonian}

In this Appendix we derive the diagonalized form of the
LL-Hamiltonian (\ref{LL Hamiltonian})  including terms of order
$1/L$, which describe integer charge and spin excitations. For
simplicity, we consider only one LL and will therefore suppress
the subscript $n$ for the leads. We start with the exact
bosonization dictionary for the Fermi-operator for electrons on
branch $r=\pm$ \cite{Haldane,Heidenreich},
\begin{eqnarray}
\psi_{rs}(x)&=&\lim\limits_{\alpha\to
0}\frac{U_{r,s}}{\sqrt{2\pi\alpha}}\exp\Big\{ir(p_{F}-\pi/L)x\Big.\nonumber\\
&+&\Big.\frac{ir}{\sqrt{2}}\Big(\phi_{\rho}(x)+s\phi_{\sigma}(x)-r(\theta_{\rho}(x)+s\theta_{\sigma}(x))\Big)\Big\}.\nonumber\\
&& \label{bosonization2}
\end{eqnarray}
The $U_{r,s}$-operator (often denoted as Klein factor) is unitary
and decreases the number of electrons with spin $s$  on branch $r$
by one. This operator also ensures the correct anticommutation
relations for $\psi_{rs}(x)$. The normal ordered charge density
operator is
$\rho(x)=\sum_{sr}:\psi_{rs}^{\dagger}(x)\psi_{rs}(x):$, where : :
measures the corresponding quantity relative to the groundstate
with all single particle states filled up to the chemical
potential $\mu_{l}$. The  normal ordered spin density operator is
defined by
$\sigma^{z}(x)=\sum_{sr}s:\psi_{rs}^{\dagger}(x)\psi_{rs}(x):\,.$
In addition, one can define (bare) current densitiy operators  for
charge $j_{\rho}=\sum_{sr}\,r\psi_{rs}^{\dagger}(x)\psi_{rs}(x)$,
and for the spin
$j_{\sigma}=\sum_{sr}\,rs\,\psi_{rs}^{\dagger}(x)\psi_{rs}(x)$,
respectively. Note that the current density has not to be normal
ordered since its groundstate expectation value vanishes. The
normal ordered product $:\psi_{rs}^{\dagger}(x)\psi_{rs}(x):$ is
calculated according to
\begin{equation}
 :\psi_{rs}^{\dagger}(x)\psi_{rs}(x):=\lim\limits_{\Delta x\to 0}:\psi_{rs}^{\dagger}(x+\Delta x)\psi_{rs}(x):.
\label{densitycal}
\end{equation}
By expanding the operator product in (\ref{densitycal}) within the
normal-order sign, the right-hand side of (\ref{densitycal})
equals
$(1/2\pi)\partial_{x}(\phi_{\rho}(x)+s\phi_{\sigma}(x)-r(\theta_{\rho}(x)+s\theta_{\sigma}(x)))/\sqrt{2}$,
from which one easily finds
\begin{equation}
\rho(x)=\frac{\sqrt{2}}{\pi}\partial_{x}\phi_{\rho}(x),\,\,\,\,\,\,
\sigma^{z}(x)=\frac{\sqrt{2}}{\pi}\partial_{x}\phi_{\sigma}(x),
\end{equation}
and for the current densities
\begin{equation}
j_{\rho}(x)=-\sqrt{2}\,\Pi_{\rho}(x),\,\,\,\,\,\,j_{\sigma}(x)=-\sqrt{2}\,\Pi_{\sigma}(x).
\end{equation}
The field $\Pi_{\nu}(x)$ is  related to $\theta_{\nu}(x)$ by
$\partial_{x}\theta_{\nu}(x)=\pi\,\Pi_{\nu}(x)$. We decompose the
phase fields into
$\phi_{\nu}(x)=\phi_{\nu}^{P}(x)+\phi_{\nu}^{0}(x)$ and
$\Pi_{\nu}(x)=\Pi_{\nu}^{P}(x)+\Pi_{\nu}^{0}(x)$, where the part
with non-zero momentum $\phi_{\nu}^{P}$ and $\Pi_{\nu}^{P}$ can be
expanded in a series of normal modes
\begin{equation}
\phi_{\nu}^{P}(x)=\frac{1}{\sqrt{L}}\sum\limits_{p\neq
0}\,\frac{1}{\sqrt{2\omega_{p\nu}}}e^{ipx}\,e^{-\alpha
|p|/2}\,(b_{\nu p}+b_{\nu-p}^{\dagger}),
\end{equation}
and for the canonical momentum
\begin{equation}
\Pi_{\nu}^{P}(x)=\frac{-i}{\sqrt{L}}\sum\limits_{p\neq
0}\,\sqrt{\frac{\omega_{p\nu}}{2}}e^{ipx}\,e^{-\alpha
|p|/2}\,(b_{\nu p}-b_{\nu-p}^{\dagger}).
\end{equation}
These fields have to satisfy bosonic commutation relations
$[\phi_{\nu}^{P}(x),\Pi_{\mu}^{P}(x')]=i\delta_{\nu\mu}(\delta(x-x')-1/L)$,
which in turn demands $[b_{\nu p},b_{\mu
p'}^{\dagger}]=\delta_{\nu\mu}\delta_{pp'}$ and $[b_{\nu p},b_{\mu
p'}]=[b_{\nu p}^{\dagger},b_{\mu p'}^{\dagger}]=0$. The zero mode
parts $\phi_{\nu}^{0}$ and $\Pi_{\nu}^{0}$ can be found by
considering the integrated charge (spin) and charge-
(spin)-currents, respectively. For instance the integrated charge
density $\sum_{rs}N_{rs}=\int
dx\rho(x)=(\sqrt{2}/\pi)(\phi_{\rho}(L/2)-\phi_{\rho}(-L/2))=(\sqrt{2}/\pi)(\phi_{\rho}^{0}(L/2)-\phi_{\rho}^{0}(-L/2))$.
Similar results are obtained for the other density operators,
which then implies the zero modes to be
$\phi_{\nu}^{0}(x)=(\pi/L)(N_{+\nu}+N_{-,\nu})x$ and
$\Pi_{\nu}^{0}=-(1/L)(N_{+\nu}-N_{-,\nu})$, where
$N_{r\rho/\sigma}=(N_{r\uparrow}\pm N_{r\downarrow})/\sqrt{2}$.
The LL-Hamiltonian (\ref{LL Hamiltonian}) is then diagonalized by
the following expansion of the bosonic fields
\begin{eqnarray}
\phi_{\nu}(x)&=&\sum\limits_{p\neq 0}\sqrt{\frac{\pi
K_{\nu}}{2|p|L}}\,e^{ipx}\,e^{-\alpha|p|/2}\,(b_{\nu p}+b_{\nu
-p}^{\dagger})\nonumber\\ &+&\frac{\pi}{L}(N_{+\nu}+N_{-,\nu})x,
\label{fullphi}
\end{eqnarray}
and for the canonical conjugate momentum operator,
\begin{eqnarray}
\Pi_{\nu}(x)&=&-i\sum\limits_{p\neq 0}\sqrt{\frac{|p|}{2\pi L
K_{\nu}}}\,e^{ipx}\,e^{-\alpha|p|/2}\,(b_{\nu p}-b_{\nu
-p}^{\dagger})\nonumber\\ &-&\frac{1}{L}(N_{+\nu}-N_{-,\nu}),
\label{fullpi}
\end{eqnarray}
where we have used $\omega_{p\nu}=|p|/K_{\nu}\pi$.  For the
operator\\
 $K_{L}=H_{L}-\mu_{l}N$ we then obtain
\begin{eqnarray}
K_{L}&=&\sum\limits_{p\neq
0,\nu}\,u_{\nu}|p|\,b_{p\nu}^{\dagger}b_{p\nu}\nonumber\\
&+&\frac{2\pi}{L}\left[\frac{u_{\nu}}{K_{\nu}}(N_{+\nu}+N_{-,\nu})^{2}+u_{\nu}K_{\nu}(N_{+\nu}-N_{-,\nu})^{2}\right].\nonumber\\
&& \label{fullH}
\end{eqnarray}
In (\ref{fullH}) we have subtracted the zero point energy
$(1/2)\sum_{p\neq 0,\nu}u_{\nu}|p|$, which originates from an
infinite filled Dirac sea of negative energy particle states in
the LL-model. The zero modes in (\ref{fullphi}) and (\ref{fullpi})
give rise to contributions of order $1/L$ in the Hamiltonian
(\ref{fullH}), and they are also responsible for a shift of the
Fermi wavevector $p_{F}$, appearing in the Fermionic field
operator $\psi_{ns}(x)$, by a contribution of the same order.
Since we are only interested in the thermodynamic limit, we have
neglected  the zero mode contributions in explicit calculations in
the main text.

\section{Exact Results for the Time Integrals}

In this Appendix we give the exact results for the time integrals in (\ref{Int1}) and (\ref{currentI2}). The integrals over the time variable $t$ appearing in (\ref{Int1}) and (\ref{currentI2}) have the form
\begin{widetext}
\begin{equation}
\int\limits_{-\infty}^{\infty}dt\,\frac{e^{i2\mu t}}{\left(\frac{\Lambda}{u_{\rho}}+it\right)^{Q}\left(\frac{\Lambda}{u_{\sigma}}+it\right)^{R}}
=\frac{2\pi e^{-2\mu\frac{\Lambda}{u_{\rho}}}\left(2\mu\right)^{Q+R-1}}{\Gamma (Q+R)}\,\,_{1}F_{1}\left(R;\,Q+R;\,2\Lambda(u_{\rho}^{-1}-u_{\sigma}^{-1})\mu\right),
\label{exact1}
\end{equation}
\end{widetext}
with $Q,R\ge 1$, but in general this integral is valid for $Q,R$ satisfying Re$(Q+R)>1$ (see \cite{G/R} p.\,345). The function $\Gamma(x)$ in (\ref{exact1}) is the Gammafunction and $_{1}F_{1}(\alpha;\,\gamma;\,z)$ is the confluent hypergeometric function given by
\begin{equation}
_{1}F_{1}(\alpha;\,\gamma;\,z)=1+\frac{\alpha}{\gamma}\frac{z}{1!}+\frac{\alpha(\alpha+1)}{\gamma(\gamma+1)}\frac{z^{2}}{2!}+\cdots\,\,\,.
\end{equation}
In the main text we considered only the leading order term of $_{1}F_{1}$ since higher order terms are small by the parameters $2\mu\Lambda/u_{\rho}$ and $2\mu\Lambda/u_{\sigma}$. 
The integrals over the delay times $t'$ and $t''$ in (\ref{currentI2}) contain Hankelfunctions of the first and second kind which are linear combinations of Besselfunctions of the first and second kind, i.e.  ${\rm H_{0}^{(1/2)}}(t\Delta)=J_{0}(t\Delta)\pm iY_{0}(t\Delta)$. The inegrals over $t'$ and $t''$ in (\ref{currentI2}) are therefore linear combinations of integrals of the form 
\begin{widetext}
\begin{equation}
\lim\limits_{\eta\to 0}\,\int\limits_{0}^{\infty}dt\,e^{-\eta t}\,Y_{0}(t \Delta)\,t^{\delta}
=\lim\limits_{\eta\to 0}\left(-\frac{2}{\pi}\Gamma(\delta +1)(\Delta^{2}+\eta^{2})^{-\frac{1}{2}(\delta +1)}\,Q_{\delta}\left(\eta/\sqrt{\eta^{2}+\Delta^{2}}\right)\right),
\end{equation}
\end{widetext}
and 
\begin{widetext}
\begin{equation}
\lim\limits_{\eta\to 0}\,\int\limits_{0}^{\infty}dt\,e^{-\eta t}\,J_{0}(t \Delta)\,t^{\delta}
=\lim\limits_{\eta\to 0}\left(\Gamma(\delta +1)(\Delta^{2}+\eta^{2})^{-\frac{1}{2}(\delta +1)}\,P_{\delta}\left(\eta/\sqrt{\eta^{2}+\Delta^{2}}\right)\right).
\end{equation}
\end{widetext}
This result is valid for $\delta>-1$ (see \cite{G/R} p. 691). The functions Q and P are Legendre functions. The limit $\eta\rightarrow 0$ for $Q_{\delta}(\eta/\sqrt{\eta^{2}+\Delta^{2}})$ is (see \cite{G/R} p. 959)
\begin{equation}
Q_{\delta}(0)=-\frac{1}{2}\sqrt{\pi}\sin(\delta\pi/2)\,\frac{\Gamma\left(\frac{\delta+1}{2}\right)}{\Gamma\left(\frac{\delta}{2}+1\right)},
\end{equation}
and the limit $\eta\rightarrow 0$ for $P_{\delta}(\eta/\sqrt{\eta^{2}+\Delta^{2}})$ is 
\begin{eqnarray}
P_{\delta}(0)&=&\frac{\sqrt{\pi}}{\Gamma\left(\frac{\delta}{2}+1\right)\Gamma\left(\frac{1-\delta}{2}\right)}\nonumber\\
&=&\frac{1}{\sqrt{\pi}}\cos(\delta\pi/2)\,\frac{\Gamma\left(\frac{\delta+1}{2}\right)}{\Gamma\left(\frac{\delta}{2}+1\right)}.
\end{eqnarray}


\begin{references}
\bibitem{note}
That superconducting correlations can induce non-local
spin-entanglement was pointed out already in \cite{CBL}. There two
superconductors, held at the same chemical potential are
tunnel-coupled to two spatially separated quantum dots containing
one electron-spin each and which are not tunnel-coupled among
themselves. The tunnel-coupling to the superconductors then induce
an effective Heisenberg spin interaction between the two spins on
the dots.
\bibitem{Kanenote}
In context with nanotubes we remark that it was pointed out
\cite{Egger,Kane} that for the (metallic) armchair nanotubes
$(N,N)$ with $N\stackrel{>}{\sim}10$ it is appropriate to use a
Luttinger liquid model where backscattering and umklapp scattering
can be neglected. This is so because for large $N$, the
probability for two electrons to be near each other is small
($\sim 1/N$) and therefore one can neglect the short range part
$r\sim a$, with $a$ the lattice spacing, of the Coulomb potential
which corresponds to large momentum transfers $\sim 2p_{F}$ at
which backscattering and umklapp scattering takes place.
\end{references}
\end{document}